\documentclass[11pt]{article}
\usepackage{amsmath,amssymb,textcomp,gensymb} 	
\usepackage[range-phrase = --]{siunitx} 				
\usepackage{booktabs, colortbl} 					
\usepackage{float} 								
\usepackage{fullpage} 							
\usepackage{caption}							
\usepackage[activate={true,nocompatibility},final,tracking=true,kerning=true,spacing=true,factor=1100,stretch=10,shrink=10]{microtype} 

\usepackage{cite} 										
\usepackage[comma, numbers, sort&compress, super]{natbib}		
\usepackage[pdfborder={0 0 0}, colorlinks, citecolor=blue]{hyperref} 	
\usepackage{rotating}									

\makeatletter
\newcommand{\mytilde}{\raise.17ex\hbox{$\scriptstyle\sim$}} 		
\renewcommand\@biblabel[1]{#1.} 							

\newcommand{\spacing}[1]{\renewcommand{\baselinestretch}{#1}\large\normalsize}
\spacing{1}

\def\@maketitle{\newpage\spacing{1}\setlength{\parskip}{12pt}%
    {\Large\bfseries\noindent\sloppy \textsf{\@title} \par}%
    {\noindent\sloppy \@author}%
}

\newenvironment{affiliations}{%
    \setcounter{enumi}{1}%
    \setlength{\parindent}{0in}%
    \slshape\sloppy%
    \begin{list}{\upshape$^{\arabic{enumi}}$}{%
        \usecounter{enumi}%
        \setlength{\leftmargin}{0in}%
        \setlength{\topsep}{0in}%
        \setlength{\labelsep}{0in}%
        \setlength{\labelwidth}{0in}%
        \setlength{\listparindent}{0in}%
        \setlength{\itemsep}{0ex}%
        \setlength{\parsep}{0in}%
        }
    }{\end{list}\par\vspace{12pt}}

\renewenvironment{abstract}{%
    \setlength{\parindent}{0in}%
    \setlength{\parskip}{0in}%
    \bfseries%
    }{\par\vspace{-6pt}}

\makeatother

\title{Molecular Tuning of the Magnetic Response in Organic Semiconductors}

\author{Erik R.~McNellis$^1,*$, Sam Schott$^2$, Henning Sirringhaus$^2$, Jairo Sinova$^1$}

\begin{document}

\maketitle

\begin{affiliations}
\item{INSPIRE Group, Johannes Gutenberg University, Staudingerweg 7, D-55128 Mainz, Germany}
\item{Cavendish Laboratory, University of Cambridge, Cambridge CB3 0HE, UK}
\\ \textrm{ * emcnelli@uni-mainz.de}
\end{affiliations}

\begin{abstract}
The tunability of high-mobility organic semi-conductors (OSCs) holds great promise for molecular spintronics. In this study, we show this extreme variability - and therefore potential tunability - of the molecular gyromagnetic coupling ("{\em g}-") tensor with respect to the geometric and electronic structure in a much studied class of OSCs. Composed of a structural theme of phenyl- and chalcogenophene (group XVI element containing, five-membered) rings and alkyl functional groups, this class forms the basis of several intensely studied high-mobility polymers and molecular OSCs. 
We show how in this class the {\em g}-tensor shifts, $\Delta g$, are determined by the effective molecular spin-orbit coupling (SOC), defined by the overlap of the atomic spin-density and the heavy atoms in the polymers. We explain the dramatic variations in SOC with molecular geometry, chemical composition, functionalization, and charge life-time using a  first-principles theoretical model based on atomic spin populations. Our approach gives a guide to  tuning the magnetic response of these OSCs by chemical synthesis.
\end{abstract}

\section{\label{sec:intro}Introduction}

From a materials science point of view, a key goal of spintronics is the purposeful tuning of the interaction of electronic spins present in the material with their motion and with  magnetic fields. In this work we
use first-principles theory to guide the engineering of the gyromagnetic
coupling ("{\em g}-") tensor in a class of high-mobility organic semiconductors. In these molecules, the {\em g}-tensor depends almost exclusively on the molecular spin-orbit coupling (SOC), which in turn is directly dependent on the molecular spin density distribution. The latter, and by extension the {\em g}-tensor, can be tuned via chemical composition and -substitution, molecular geometry and -functionalization.

As in related efforts in molecular electronics,\cite{Carroll2002} photonics and photovoltaics,\cite{Hains2010} the expected benefits of crafting spintronic devices from molecules\cite{Koopmans2014} include reduced cost of production, increased material tailorability and -abundance, and disruptive technologies outperforming traditional components in given areas, such as OLEDs.
Molecular spintronics holds great potential as a complement to traditional
materials, in long spin life-time, hybrid organic-inorganic designs,\cite{Dediu2009} and
is central to the concept of using molecules to tune solid interfaces
for spintronic applications, so-called 'spinterfaces'.\cite{Sanvito2010,Galbiati2014}
While the fundamental physics of spintronic phenomena are the same in
molecules as in the solid state, routes to achieving similar material
properties can be very different.

Electronic spins are modulated by magnetic fields, and the SOC, or coupling between spin and orbital angular momenta. Spintronic materials design therefore chiefly amounts to tuning magnetic fields,
the coupling of the spin to those fields, and the effective spin-orbit
coupling in terms of strength and spatial distribution. In molecular
spintronics, this in turn translates to synthesizing molecules of
optimal hyperfine field and molecular SOC, in addition to general
requirements on composition and structure. As an aspect of the electronic
structure, the {\em g}-tensor depends directly on the molecular SOC.

In a recent publication,\cite{Schott2017} we have studied the relationship
between the molecular spin density, SOC, and \emph{g}-tensors in a
much studied class of high-mobility organic semi-conductors (OSCs).
While our results show these properties to be sensitive to molecular composition
and structure to a degree running counter to established chemical
intuition, they also highlight the great potential for purposeful
tuning of molecular spintronic components, in particular when guided
by fairly straightforward simulations from first-principles theory.
We also point out the influence of the \emph{effective} molecular
SOC and \emph{g}-tensor shifts from the free electron value in this
range of molecules. In this work we significantly expand on
the first-principles theoretical model previously used, further raising
the quality of obtained predictions, extend our class of molecules,
address charge life-time effects, and elaborate on methodological
aspects.

\begin{figure}
\includegraphics[width=\columnwidth]{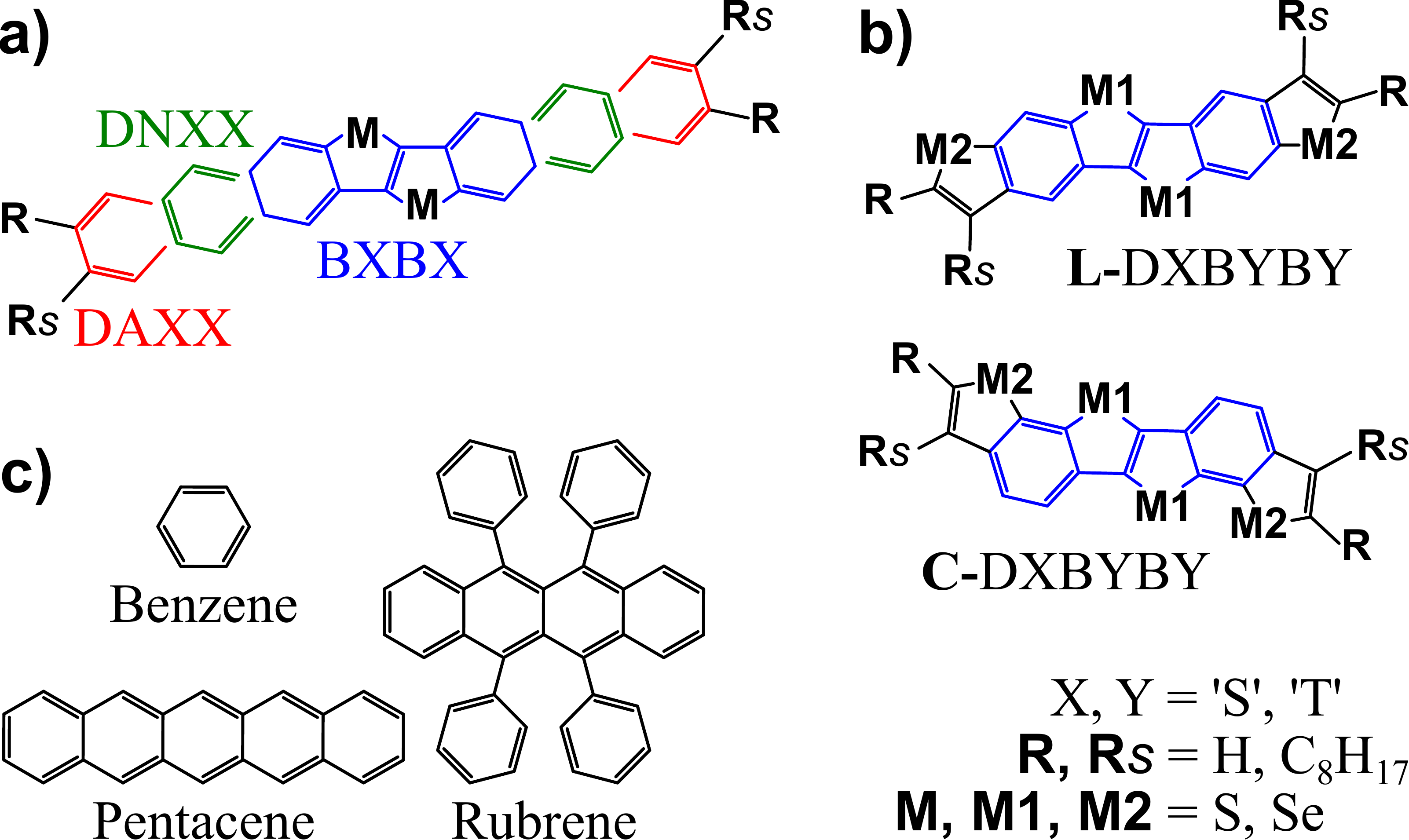}
\caption{\label{fig:fig1} Structural formulae of all molecules studied in this work. a) Single chalcogen pair molecules, with the central BXBX (X = 'T' and 'S' for M = sulfur and selenium, respectively) structure shown in blue. Adding one and two sets of fused phenyl rings gives DNXX (green) and DAXX (red), respectively. Further shown is the bonding site of the C8 alkyl chain functional group (R), and that same group shifted to the adjacent bonding site (R{\em s}). b) Linear (L) and curved (C) dual chalcogen pair molecules DXBYBY, including the mixed pair DSBTBT molecules, where M1 = sulfur and M2 = selenium. Otherwise with nomenclature and functionalization as in a). c) The pure hydrocarbons benzene, rubrene and pentacene included for improved fit quality.}
\end{figure}

We focus on a class of chalcogenophene (five-membered rings composed of carbon and a single group XVI atom) based OSCs with an [1]benzothieno[3,2-b][1]-benzothiophene (BTBT) molecule as the central structural element (see Fig.~\ref{fig:fig1}). BTBT consists of two fused thiophene ($\mathrm{C_{4}S}$) rings extended by fused phenyl ($\mathrm{C_{6}H_{4}}$) moieties on opposite
sides. While BTBT and its synthesis has been known since 1949,\cite{WesleyHorton1949}
its potential as a high-mobility ($\mathrm{\mu>1.0\:[cm^{2}/Vs]}$)
OSC has only recently been realized. The basic structural theme of
a conjugated molecule consisting of one or several chalcogenophene-
and phenyl-rings, functionalized with alkyl chains for improved solubility
and thin-film morphology, has come to form a class of high mobility
OSCs.\cite{Takimiya2014}

Compositional variations in this class include substitutions with heavier chalcogens, functionalization with alkyl chains, and extension of the conjugation length by adding fused phenyl rings. Substituting the thiophene rings in BTBT for selenophene produces the 'BSBS' analogue.\cite{Takimiya2006a} Alkylation is generally done to improve solubility and thin-film morphology, with molecules with {\em n}-membered chains labeled C{\em n}-X (e.g. 'C8-BTBT').\cite{Ebata2007,Izawa2008} Substituting the single phenyl ring ('benzo-') moieties in BTBT for double ring ('dinaphta-') moieties yields the so-called DNTT molecule.\cite{Yamamoto2007} The same molecule with a three ring ('dianthra-') moiety is called DATT.\cite{Takimiya2014} The Se-substituted analogues of DNTT and DATT are labeled DNSS and DASS, respectively. Some of these compounds have been employed as high-mobility organic field-effect and thin-film transistors (OFETs and OTFTs).\cite{Izawa2008,Matsui2012}

However, minor variations in composition can also produce significantly lower hole mobilities,
which has been attributed to the large differences in HOMO orbital
weight on the chalcogen atoms observed in first-principles calculations.\cite{Takimiya2006a}
This aligns very well with our own finding that the tunability of
magnetic response with structural variations of these molecules can
be attributed to the corresponding variation of spin density weight
at the chalcogen atoms.\cite{Schott2017}

Notably, chalcogenophenes and alkyl functional groups also form the
basis of many intensively studied high-mobility polymers, such as
PBTTT\cite{McCulloch2006} and P3HT.\cite{Heeger2010} While these
remain outside the scope of this work, we expect our key insights
to also apply to such polymers to some degree.

This paper is organized as follows: In section \ref{sec:methods} we describe the set of molecules studied in greater detail, followed by a presentation of a simple model devised to describe the relationship between {\em g}-tensor shifts and the molecular spin density distribution. This model is parametrized by electronic structure calculations from first-principles theory, the methodological detail of which is explained in the following subsection. All first-principles results, as well as the quality and applicability of our {\em g}-tensor shift model, are presented and discussed in context in section \ref{sec:res}, and finally summarized in section \ref{sec:summary}.

\section{Studied Molecules and Methodology}
\label{sec:methods}

\subsection{Target Class of OSC Molecules}

The set of chalcogenophene OSCs studied here is illustrated using structural formulae in Fig.~\ref{fig:fig1}.  The BXBX molecule is shown in blue in Fig.~\ref{fig:fig1}a (X = 'T', 'S' for sulfur and selenium chalcogens, respectively), along with it's extension of the fused phenyl moieties to DNXX and DAXX, forming a total of three basic chalcogenophene structures with a single chalcogen atom pair. Further shown is the position of the eight-membered alkyl chain ($\mathrm{C_{8}H_{17}-}$) functional group {\bf R}. This is generally how these molecules have been alkylated in literature.\cite{Takimiya2014,Ebata2007,Izawa2008} However, as will be shown below, the spin density weight on the conjugated moiety at the alkyl chain bonding site modifies the influence of the functional group on the {\em g}-tensor shift. In most of the molecules studied here, there is a spin density maximum on the conjugated moiety at the bonding site {\bf R} in Fig.~\ref{fig:fig1}, and a minimum at the adjacent, "shifted" site of {\bf R}{\em s}. This is illustrated in Fig.~\ref{fig:fig2}. Consequentially, we also include molecules functionalized at the shifted bonding site in our target class.

All combinations of the three single chalcogen pair molecules with the three possibilities of no alkyl chain, alkylation at the ordinary and shifted sites amount to nine sulfur-based molecules, which with their selenium-substituted analogues add up to 18 molecules.

The next variation of the BXBX structural motif is shown in Fig.~\ref{fig:fig1}b - the incorporation of a second set of chalcogenophene rings. These can be added forming a linear shape with e.g.~the BTBT molecule, which we denote L-DTBTBT ("dibenzothiopheno[6,5-b:6',5'-f]thieno[3,2-b]thiophene"), or forming a curved 'S' shape ("dibenzothiopheno[7,6-b:7',6'-f]thieno[3,2-b]thiophene"), which we correspondingly label C-DTBTBT. We include all variations of selenium substitutions (e. g. 'L-DSBSBS'), no alkylation and alkylation at the two possible sites, for a total of twelve distinct molecules added to the preceding 18.

Additionally, as a way of cross-referencing the sulfur and selenium fits, we add the latter set with an inner sulfur-, and outer selenium-atom pair, e. g. L-DSBTBT, which with the curved and linear forms and the three alkylation options add a further six molecules to the 30 already described.

Finally, in order to improve the quality of the fitted carbon atom coefficients, the chalcogen-free hydrocarbons benzene, pentacene and rubrene are included in the fit, for a total of 39 distinct molecules. Structural formulae of the latter are shown in Fig.~\ref{fig:fig1}c.

These OSCs are non-magnetic, and in a real material or experimental sample only acquire a spin (and therefore, a {\em g}-tensor) when charged. Depending on the nature of the material, these charges may be short- or long-lived, such as in a close-packed semi-crystal or in solution, respectively. Since actual material models are outside the scope of this study, we perform calculations on the 39 molecules in geometries optimized with respect to both a charge neutral (0) and a singly positively charged (+1), cationic state. This simulates the structural effect on the {\em g}-tensor in the limits of zero and infinite charge life-time, respectively.

The neutral benzene molecule has $\mathrm{D_{6h}}$ point-group symmetry. Cationic benzene, and both charge states of rubrene and pentacene here have $\mathrm{D_{2h}}$ symmetry. All chalcogenophenes have $\mathrm{C_{2h}}$ point-group symmetry in both charge states.

\subsection{Ansatz for Molecular SOC: {\em g}-Tensors Linearly Dependent on Atomic Spins}
\label{sub:linmodel}

We begin by briefly summarizing the {\em g-}tensor theory underpinning
this work. Following Stone\cite{Stone1963} in the Neese formulation,\cite{Neese2003}
the gauge-invariant molecular {\em g-}tensor can be written as

\begin{equation}
\mathbf{g=g_{e}+\Delta g^{RMC}+\Delta g^{GC}+\Delta g^{OZ/SOC}}\,,\label{eq:g}
\end{equation}
where $\mathbf{g_{e}}$ is the Land{\'e} {\em g-}factor of free electrons (here treated as a scaling factor for the rank three unit tensor, for consistency), and
the latter three terms form the shift tensor $\mathbf{\Delta g}$ from the free electron
value.  The isotropic shift $\Delta g$ is obtained by averaging over the Cartesian components of $\mathbf{\Delta g}$. The first two terms of Eq.~\ref{eq:g} are a relativistic mass correction
and gauge correction, respectively, and the last is a mixed second
derivative of the total energy, with contributions from the orbital
Zeeman and SOC terms.

The relative magnitudes of these three terms in the studied class of molecules are a crucial point in our reasoning. Relativistic effects, to include mass corrections, are generally small in organic molecules. If also the gauge correction is sufficiently small, the molecular \emph{g-}tensor shifts $\Delta g$ are dominated by $\Delta g^{\mathrm{OZ/SOC}}$, in turn making the molecular SOC the tuning parameter determining $\Delta g$. As will be shown in section \ref{sec:res}, this is indeed the case for our target molecules. A stronger influence of the first two terms is however perfectly possible, depending on the molecular chemical composition. In practice, relative magnitudes of the $\Delta g$ terms must be heuristically determined.

$\Delta g^{\mathrm{OZ/SOC}}$ is linear in the molecular SOC, which is dominated by one-electron scalar products between electronic spins and molecular orbital angular momentum. Many-body terms such as the
two-body spin-spin-orbit (SSO) and spin-other-orbit (SOO) interactions also contribute, but to a significantly lesser degree. Since SOC is weak on the scale of other electronic interactions, and the two-body SOC terms therefore usually negligible, SOC is often modeled as a local operator in solid state systems. If its spin density can be sufficiently well approximated as a sum of atomic contributions, this approximation holds also for a molecule. In order to compare differences in molecular SOC between molecules of differing geometry and composition, we therefore make the Ansatz that $\Delta g^{\mathrm{OZ/SOC}}$ is linear in the molecular spin density, approximated as a sum of localized atomic spins, with
element-dependent proportionality constants representing the net orbital angular momentum interaction.

That is, we make the approximation 

\begin{equation}
\Delta g^{\mathrm{OZ/SOC}}\approx\sum_{e=1}^{M}c_{e}\sum_{i=1}^{N}\sigma_{i}^{e}\,,\label{eq:linmodel}
\end{equation}
where $\sigma_{i}^{e}$ is the effective spin at atom $i$ of element $e$, $N$ is the number of atoms of element $e$ in the molecule, $M$ is the number of different elements and $c_{e}$ a constant.
While such localized approximations of SOC are common in solid-state models,\cite{AshcroftMermin} their utility for molecules
is considerably less obvious. The key benefit of this approximation is that it casts changes in the (here) dominant part of the molecular
\emph{g}-tensor shift in terms of changes in the molecular spin density, which is readily obtained.

We cannot expect universal transferability of such an approximation. However, if our Ansatz holds sufficiently well within a given class of molecules, we may fit common $c_{e}$ for that class, and analyze
internal variations in terms of the spin density distribution (magnitude of atomic spins). In
the following, we for a target molecule approximate $\Delta g^{\mathrm{OZ/SOC}}$
by calculating atomic spins $\sigma_{i}^{e}$ for a set of molecules
including the target molecule, fitting $c_{e}$ coefficients to $\Delta g^{\mathrm{OZ/SOC}}$
terms calculated from first-principles for the set excluding the target
molecule, and evaluating Eq.~\ref{eq:linmodel} with the target $\sigma_{i}^{e}$.
This way, the target molecule is never part of its own fitting set.

Since a negative $c_{e}$ lacks physical interpretation, we use a
non-negative multivariate least-squares fitting algorithm implemented
in Scientific Python\cite{scipy0161} (version 0.16.1, using subroutine 'scipy.optimize.nnls'). Since their maximal atomic orbital momentum is zero, hydrogen atoms were excluded from the $c_{e}$ fits. As a simple but commonplace statistical test, we for each fit calculate the coefficient of determination or $R^{2}$-value according to

\begin{equation}
R^{2}\equiv1-\frac{SS_{res}}{SS_{tot}}=1-\frac{\sum_{i}(x_{i}-y_{i})^{2}}{\sum_{i}(x_{i}-\overline{x})^{2}}\,,\label{eq:r2}
\end{equation}
where $SS_{res}$, $SS_{tot}$, $x_{i}$ and $y_{i}$ are the residual
sum of squares, the total sum of squares, and the calculated and fitted
$\Delta g^{\mathrm{OZ/SOC}}$ value, respectively.

\subsection{DFT Calculations: Geometries, \emph{g-}Tensors, Spin Densities}

We use density functional theory (DFT) to calculate molecular geometries
and \emph{g}-tensors and atomic spin densities. Describing SOC effects
in organic molecules from first-principles theory is challenging,
since they are generally small but often not negligible. The standard
approach of spin-orbit DFT (SODFT) - describing SOC as a correction
to a scalar relativistic effective core potential (ECP)\cite{FernandezPacios1985}
- is well justified for heavy atoms, with deep cores decoupled from
the chemical bonding of the valence electrons, and interacting weakly
with the chemical environment.

However, in lighter elements, the frozen-core approximation of an
ECP is much more spurious, both in terms of SOC and other electronic
interactions. We therefore opt for an all-electron SODFT treatment,
with nuclear relativistic effects described by the zeroth order regular
approximation\cite{Lenthe1993a} (ZORA) with the standard point-charge
approximation of atomic nuclei. We chose the high-quality SARC\cite{Pantazis2008}
basis set family, which has been recontracted for the ZORA approximation,
testing singly and doubly polarized valence sets from single- (SVP)
to quadruple-zeta (QZVP / QZVPP) sizes. In so doing we were forced
to remove the two diffuse (augmentation) functions on the carbon atoms
in order to eliminate linear dependencies, but no other modifications
to the basis sets were made. Geometries were found fully converged
with respect to basis set size at triple-zeta (TZVP) level, but \emph{g}-tensor
shifts increased slightly ($<300$ [ppm]) at quadruple-zeta level,
worsening the comparison to experimental measurements, which indicates
that the quality of TZVP \emph{g}-tensor shift predictions is partly
due to cancellation between basis set and other errors.

Deficiencies due to electron delocalization error\cite{Cohen2008} in (semi-) local exchange-correlation
(xc) DFT functionals are particularly severe for molecules\cite{Ernzerhof1999}
and magnetic phenomena. Therefore, hybrid xc-functionals, with non-local
exact exchange added to the (semi-)local terms, perform better for
the calculation of {\em g}-tensors.\cite{Wilson2005} All calculations
presented here were performed using the PBE0\cite{Adamo1998} hybrid
xc-functional, which has been shown\cite{Neese2001} to perform excellently for similar systems, including almost reproducing\cite{Fonari2014} the $G_{0}W_{0}$ bandstructure of pentacene and rubrene.

All calculations were performed on single molecules, approximating molecules dissolved in solution. All {\em g}-tensor calculations
were performed on positively charged (cationic) molecules. For each
molecule, the geometry was obtained by unrestricted geometry optimization
in the charge neutral and cationic states, simulating the limits of
short- and long life time of the charged state, respectively. All
geometry optimizations were carried out using the NWChem quantum chemistry
software, version 6.5.\cite{Valiev2010}

\emph{g}-tensors of cationic molecules were calculated using optimized
geometries at the exact same level of theory, using the coupled-perturbed
Kohn-Sham technique\cite{Neese2001} and SOC operator approximation\cite{Neese2005}
developed by Neese, and implemented in the ORCA software package,\cite{Neese2012}
version 3.0.3.

Calculating atomic spins from molecular spin densities in a consistent
and transferable manner is difficult, since it requires a solution
to the atoms-in-molecules (AIM) problem\cite{Bader1991}. While the
AIM problem cannot be solved rigorously, several levels of approximations
exist. The most readily available such method is the assignment of
atomic spin from a Mulliken\cite{Mulliken1955} or L{\"o}wdin\cite{Lowdin1950}
population analysis. These techniques partition a molecular wavefunction,
described by a linear combination of atomic orbital (LCAO) basis functions,
into atomic contributions based on which atom each basis function
is centered on. That makes Mulliken or L{\"o}wdin partial charges strongly
basis set dependent, and by way of the basis set also dependent on
the geometry. Furthermore, such a partitioning of charge density is
highly ambiguous in the interstitial region between atoms in a molecule,
where basis functions from neighboring atoms overlap. This spuriousness
leads to unphysical, inconsistent variations in the calculated atomic
spins,\cite{Philips2010} which in turn causes large statistical scatter
in a fit on the form of Eq.~\ref{eq:linmodel} as described above,
and a tendency for an unrestricted fitting algorithm to produce negative
$c_{e}$ coefficients.

A more rigorous approach, based on the partitioning of charge density
by surfaces where the charge density is stationary in space, is so-called
Bader partitioning.\cite{Bader1985} This method, like the calculated
molecular charge density it takes as input, is not basis set dependent
when converged with respect to the basis set. The Bader method significantly
improved the fit results over Mulliken or L{\"o}wdin population analysis,
with further minor improvements when using an improved grid integration
method.\cite{Yu2011a} The highest quality fits, characterized by
small statistical scatter, numerically stable fits, and consistently
positive $c_{e}$ coefficients were obtained with a method\cite{FonsecaGuerra2004}
partitioning the Voronoi deformation density (VDD), however. All of the
density partitioning AIM methods were employed using the 'Bader' program
(version 0.95a) developed by Henkelman et al.\cite{Tang2009a}

In summary, the calculations were carried out as follows: for each
molecule, the fully optimized geometry was calculated for the neutral
and cationic molecule. Then the \emph{g}-tensor and molecular spin
density of the resulting geometries were obtained in separate single-point
calculations. The spin density was then partitioned into atomic spins by the 'Bader' program using the VDD method, which in turn were fed into the fitting procedure described in the previous subsection.

\section{\label{sec:res} Results and Discussion}

\begin{figure*}
\includegraphics[width=\textwidth]{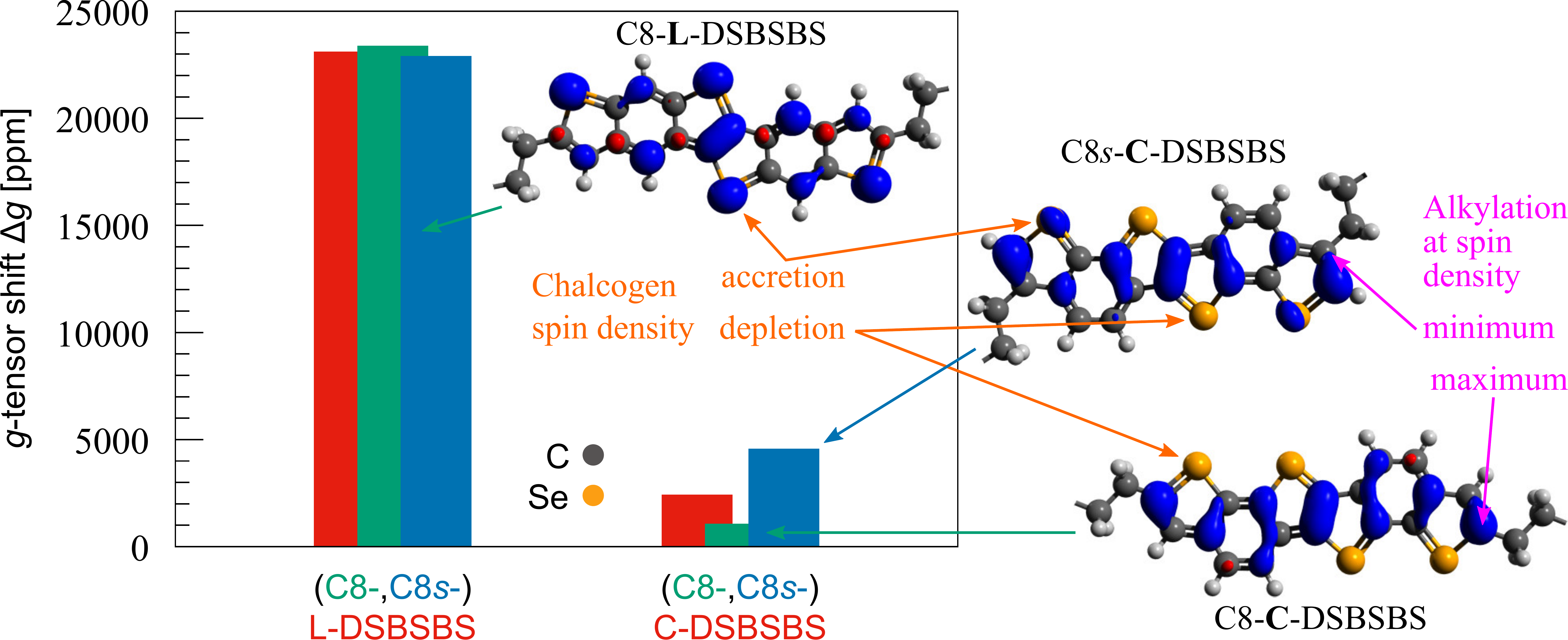}
\caption{\label{fig:fig2} Illustration of the relationship between predicted {\em g}-tensors and molecular spin density distributions, and the variations of the latter with molecular geometry and functionalization. At left (right) the histogram shows calculated $\Delta g$ for the linear (curved) DSBSBS molecules without alkyl chains, and C8 / C8{\em s} alkyl chains bonded at spin density maxima and -minima. As shown by the blue isocontour in the accompanying spin density plots, the L-DSBSBS molecule has heavy spin density weight at the chalcogen atoms - $\Delta g$ is large. The spin density structure is not significantly changed when alkylating L-DSBSBS at the two sites. The {\em opposite} is true in the C-DSBSBS molecule, {\em despite its identical chemical composition}: alkylating C-DSBSBS at a bonding site corresponding to a spin density maximum (C8-C-DSBSBS) pulls spin density away from all chalcogens, whereas alkylating at a spin density minimum (C8{\em s}-C-DSBSBS) leaves significant weight on the outer chalcogen pair. Qualitatively similar variations are found in all other molecules studied here.}
\end{figure*}

\begin{table}
\begin{centering}
\begin{tabular}{r|r|r|r|r}
\hline 
\hline
{\bf Molecule} & {\bf Exp.\cite{Schott2017}} & $\mathbf{\Delta g_{0}}$ & $\mathbf{\Delta g_{0}^{OZ/SOC}}$ & $\mathbf{\Delta g_{+1}}$-$\mathbf{\Delta g_{0}}$\\
\hline 
benzene & - & 137 & 173 & -1\\
pentacene & 311 & 352 & 346 & -7\\
rubrene & 309 & 81 & 73 & 291\\
BTBT & 2141 & 2238 & 2216 & -327\\
C8-BTBT & 1087 & 1107 & 1109 & -342\\
C8\emph{s}-BTBT & - & 2828 & 2812 & -426\\
BSBS & 10010 & 14255 & 14129 & -2502\\
C8-BSBS & 6322 & 6773 & 6696 & -3005\\
C8\emph{s}-BSBS & - & 16677 & 16548 & -2794\\
DNTT & 1959 & 2073 & 2035 & -164\\
C8-DNTT & 1657 & 1794 & 1778 & -222\\
C8\emph{s}-DNTT & - & 1769 & 1754 & -223\\
DNSS & 9772 & 10414 & 10289 & -1019\\
C8-DNSS & - & 8760 & 8662 & -1391\\
C8\emph{s}-DNSS & - & 8526 & 8429 & -1312\\
DATT & - & 1598 & 1553 & -112\\
C8-DATT & - & 1481 & 1459 & -123\\
C8\emph{s}-DATT & - & 1527 & 1506 & -107\\
DASS & - & 7031 & 6911 & -638\\
C8-DASS & - & 6415 & 6321 & -707\\
C8\emph{s}-DASS & - & 6620 & 6526 & -589\\
L-DTBTBT & - & 4176 & 4128 & -436\\
C8-L-DTBTBT & 3514 & 4180 & 4143 & -459\\
C8\emph{s}-L-DTBTBT & - & 4166 & 4105 & -350\\
L-DSBSBS & - & 23265 & 23082 & -2719\\
C8-L-DSBSBS & - & 23527 & 23353 & -3178\\
C8\emph{s}-L-DSBSBS & - & 23082 & 22882 & -2070\\
C-DTBTBT & - & 419 & 383 & -43\\
C8-C-DTBTBT & 354 & 237 & 231 & -59\\
C8\emph{s}-C-DTBTBT & - & 802 & 759 & -111\\
C-DSBSBS & - & 2498 & 2395 & -793\\
C8-C-DSBSBS & - & 1117 & 1041 & -769\\
C8\emph{s}-C-DSBSBS & - & 4672 & 4547 & -1223\\
L-DSBTBT & - & 15723 & 15619 & -1637\\
C8-L-DSBTBT & - & 16329 & 16233 & -1642\\
C8\emph{s}-L-DSBTBT & - & 17149 & 17032 & -1599\\
C-DSBTBT & - & 3274 & 3213 & -270\\
C8-C-DSBTBT & - & 1376 & 1343 & -38\\
C8\emph{s}-C-DSBTBT & - & 5297 & 5242 & -1041\\
\end{tabular}
\par\end{centering}

\caption{\label{tab:dg} Isotropic {\em g}-tensor shifts $\Delta g$ calculated using DFT for all molecules studied here, in units of [ppm]. In column 2, available experimental measurements of $\Delta g$. In columns 3 - 5, theoretically calculated {\em g}-tensor shifts $\Delta g_0$ in charge-neutral geometries, the corresponding $\Delta g_0^\mathrm{OZ/SOC}$ terms, and the change in $\Delta g$ in the cationic geometry. The cationic {\em g}-tensor shift $\Delta g_+$ can consequentially be obtained by adding columns 3 and 5.}

\end{table}

In this section all first-principles and fit results are presented and discussed in context. Subsection \ref{sub:dftg} presents all predictions of {\em g}-tensor shifts from DFT. These results are then analysed in terms of our Ansatz of a linear dependence of $\Delta g$ on the local atomic spin populations in subsection \ref{sub:linmod}. Our Ansatz is further applied and validated in the special cases of $\Delta g$ shifts upon OSC alkylation and in the long charge life-time limit. These results are presented and discussed in subsections \ref{sub:alk} and \ref{sub:clt}, respectively.

\subsection{\label{sub:dftg} DFT {\em g}-Tensor Shifts: General Trends and Comparison to Experiment}

All isotropic {\em g}-tensor shifts calculated using DFT for the 78 distinct variations of chemical composition, geometry, and charge-state geometry are presented in Table \ref{tab:dg}, in units of [ppm]. The dominance of the $\Delta g^\mathrm{OZ/SOC}$ term in the $\Delta g$ of these molecules is striking, with the RMS sum of the relativistic mass- and gauge correction terms of Eq.~\ref{eq:g} a negligible 88 [ppm]. Therefore, for all intents and purposes, shifts in $\Delta g$ of these molecules is entirely due to a shift of $\Delta g^\mathrm{OZ/SOC}$, highlighting the influence of SOC on $\Delta g$. In the following, the labels $\Delta g$ and $\Delta g^\mathrm{OZ/SOC}$ are in parts used interchangeably.

Beginning with the single chalcogen pair molecules (Fig.~\ref{fig:fig1}a), we see a) a reduction in $\Delta g$ with increasing size of the conjugated moiety, b) a change in $\Delta g$ upon alkylation of the molecules, similarly diminishing with the size of the conjugated moiety, and c) identical trends in the sulfur- and selenium based molecules, up to a roughly uniform scaling factor.

In BTBT, DNTT and DATT, the size of the conjugated moiety increases from one to two and three pairs of phenyl rings, respectively (see Fig.~\ref{fig:fig1}). This increase changes $\Delta g$, both quantitatively, in the magnitude of shifts, and qualitatively, in the effect of alkylation: the {\em g}-shift in pure BTBT is approximately 9\% greater than in pure DNTT, which in turn is 31 \% greater than in pure DATT. A visual analysis of the corresponding differences in calculated spin density similar to that of Fig.~\ref{fig:fig2} shows the spin density delocalizing over the entire conjugated moiety, and therefore concentrating less where the orbital angular momentum is greatest, at the central chalcogen atom pair. The resulting reduction in effective molecular SOC reduces the {\em g}-shift.

The effect of alkylation of these molecules also strongly depends on the extent of the conjugated system. In BTBT, the spin density is strongly confined to the conjugated moiety. Upon alkylation, spin leaks out onto the alkyl chain and depletes from the chalcogen atom pair, causing a $\sim 50\%$ reduction in the {\em g}-shift. Shifting the alkyl chain to the adjacent bonding site, corresponding to a spin density minimum, instead further concentrates spin density on the chalcogen pair, resulting in a $\sim 50\%$ {\em increase} in $\Delta g$, with the pure BTBT $\Delta g$ as a baseline. This happens despite the identical chemical composition of C8- and C8{\em s}-BTBT.

While alkylation similarly suppresses the {\em g}-shift in DNTT and DATT, the effect becomes weaker with increasing size of the conjugated moiety, and the qualitative difference between alkylation at the two sites vanishes, consistent with a weaker interaction between alkyl chain and spin density as the latter becomes less confined and more delocalized. The effect of substitution of sulfur with selenium roughly amounts to a uniform increase in magnitude of shifts, consistent with a greater orbital angular momentum but otherwise similar chemistry of the heavier chalcogen.

In the dual chalcogen pair molecules (Fig.~\ref{fig:fig1}b), the molecular geometry has a very large influence on the {\em g}-shift. See Fig.~\ref{fig:fig2}. In linear DSBSBS (L-DSBSBS), the calculated spin density shows an alternating pattern of maxima and minima (blue and red contours in Fig.~\ref{fig:fig2}, respectively), with large weight on both chalcogen atom pairs. The {\em g}-shifts are comparatively large, and alkylating the molecule at either a spin maximum or -minimum has little to no effect.

By shifting the outer chalcogenophene ring pair by one bonding site, but maintaining the exact same chemical composition, the picture changes to the opposite: in curved DSBSBS (C-DSBSBS) the pure {\em g}-shift is {\em an order of magnitude} smaller than in the L-DSBSBS, and alkylating at a spin density maximum (minimum) approximately doubles (halves) $\Delta g$. This is because the spin density of C-DSBSBS has almost no weight on the chalcogens, but heavy weight at the alkyl chain bonding site. Again, this causes spin density to leak onto the alkyl chain, reducing the spin at the chalcogens, and therefore reducing the effective molecular SOC.

Again, comparing to the corresponding sulfur-substituted molecules (L- and C-DTBTBT), we see the very same pattern, albeit reduced in magnitude, consistent with chalcogen weight appearing as a roughly uniform scaling of $\Delta g$. In the curved mixed dual chalcogen pair molecules (C-DSBTBT), the same pattern again emerges, but at absolute $\Delta g$ magnitudes between those of C-DTBTBT and C-DSBSBS - {\em g}-shifts are 'averaged' between the sulfur and selenium chalcogen pair. However, whereas the $\Delta g$ of L-DTBTBT and -DSBSBS are largely unaffected by alkylation, the $\Delta g$ of L-DSBTBT increases, particularly when alkylating at the shifted bonding site.

Comparing to the 11 already published\cite{Schott2017} experimental numbers, we note that these refer to molecules with alkyl chains 8, 10 and 12 methylene units long, whereas the theoretical structures all have 8-unit chains. Since test calculations showed no variations of {\em g}-tensor shifts beyond a chain length of about 4 units, this comparison is valid.

Theoretical predictions generally compare very well to the experimental data, with the lone exception of the BSBS molecule. The RMS error to experiment of the theoretical predictions is 342 [ppm] when excluding BSBS, which at this level of theoretical approximations can be considered negligible. When including the BSBS number, the RMS error increases to 1321 [ppm]. Charge life-time effects in the experimental numbers are possibly a contributing factor here, since the DFT prediction error of some 42 \% in the short charge life-time limit (neutral molecular geometry) shrinks to approximately 17 \% in the long charge life-time limit (cationic molecular geometry). Still, 17 \% dwarfs the error (7 \%, 451 [ppm]) in, for example, the prediction for the short charge life-time limit of C8-BSBS, for which the long charge life-time limit is a very poor approximation (38 \% error). The theoretical method used is identical for all molecules, suggesting an inconsistency in the experimental BSBS measurement not well described by the single-molecule approximation.

\subsection{\label{sub:linmod} Validation of Linear Ansatz Model}

\begin{figure}
\includegraphics[width=\columnwidth]{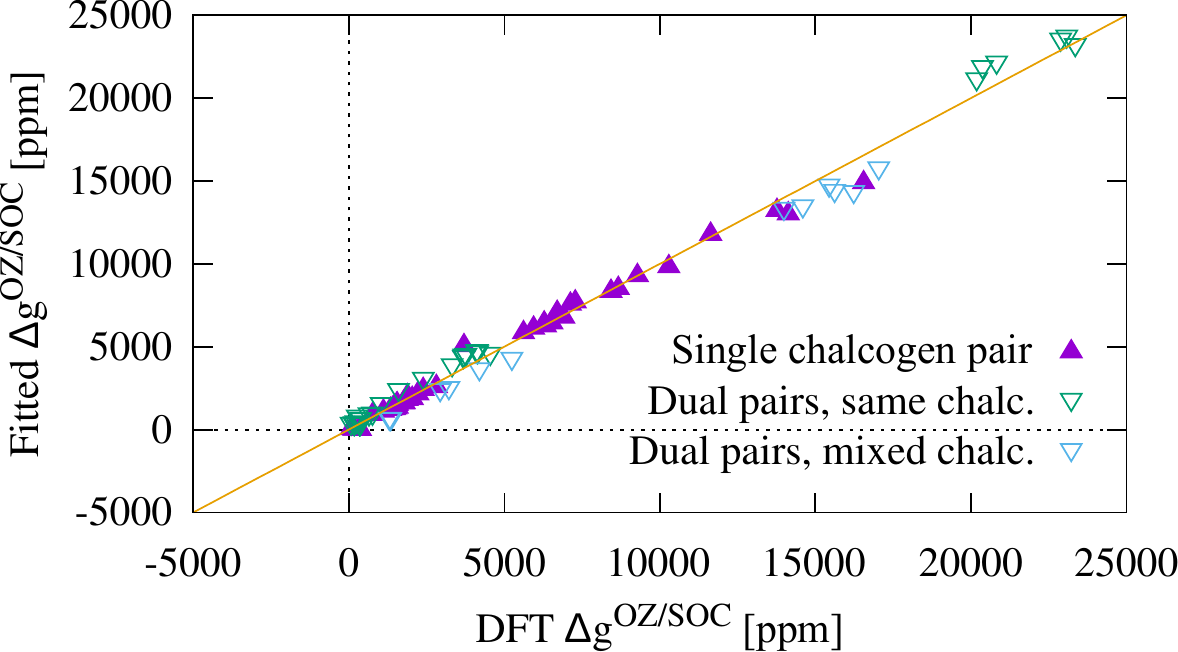}
\caption{\label{fig:fig3} Correlation plot of  $\Delta g^\mathrm{OZ/SOC}$ terms fitted on the form of Eq.~\ref{eq:linmodel}. The fit was done separately for single chalcogen pair (solid, magenta triangles) and dual chalcogen pair molecules (empty, inverted triangles). Dual pair molecules with same and mixed chalcogens were fitted together, but are colored differently to show the identical validity of the approximation regardless of chalcogen composition. The $R^2$ for the single- and dual-chalcogen pair statistics are 0.990 and 0.991, respectively.}
\end{figure}

While the picture thus far is well explained by a visual analysis in terms of differences in spin density at high orbital angular momentum atoms on the form of Fig.~\ref{fig:fig2}, a quantitative analysis on the form of Eq.~\ref{eq:linmodel} is far more exhaustive. A straight fit of all local atomic spin populations {\em versus} theoretically predicted {\em g}-shifts yields a (for a strictly linear statistical hypothesis) low $R^2$ of 0.949. Furthermore, despite using the high-quality Voronoi density partitioning method, this fit is numerically unstable for the carbon $c_e$, which comes out negative. In other words, the single fitting set of all molecules studied here is poorly described by the Ansatz of a linear dependence of $\Delta g$ on strictly localized atomic spin populations.

This is because the single-pair molecules have a single ($\mathrm{C_{2h}}$) symmetry unique chalcogen atom, but the dual-pair molecules have two. The resulting difference in chalcogen-chalcogen interactions may shift $c_e$, since it describes both one- and two-body spin-orbit interactions determining $\Delta g^\mathrm{OZ/SOC}$ as a single, local, effective coefficient. Fitting the dual-pair molecules separately yields a high $R^2$ of 0.991. This fit is shown as inverted, empty triangles in a correlation plot in Fig.~\ref{fig:fig3}. Importantly, and further underscoring the insight that geometry is the key in the variations of $\Delta g$ in the dual-pair molecules, all dual-pair molecules are fitted together. That is, the mixed chalcogen dual pair molecules (e.g.~LD{\bf S}BTBT) are described just as well by the model as the pure sulfur (e.g.~L-DTBTBT) or selenium (e.g.~L-DSBSBS) ones. Highlighting the similar performance of the linear Ansatz for all dual-pair molecules, the empty inverted triangles of same- and mixed chalcogen pair molecules are differently colored in Fig.~\ref{fig:fig3}.

The C8- and C8{\em s}-BSBS molecules produce negative carbon $c_e$ when included in the fitting set of the other single-pair molecules in the procedure described at the end of subsection \ref{sub:linmodel}. We therefore exclude these two from the fitting sets of the other molecules. With an otherwise unmodified fit procedure, the fitted single-pair molecule statistic also yields a very high $R^2$ of 0.990, which is shown as filled triangles in Fig.~\ref{fig:fig3}.

Just as with the relatively large error in $\Delta g$ predicted by DFT for BSBS, the perfect consistency of the theoretical method only allows for speculation as to why C8- and C8{\em s}-BSBS are outliers in our linear model. We conjecture that the strong confinement of the electron hole on the small BSBS moiety leads to stronger non-local interactions of the spin density when an alkyl chain is added, in violation of the local, linear dependence of $\Delta g$ underlying our quantitative Ansatz.

The average $c_e$ obtained in the fit are presented in Table \ref{tab:ce}. Since each molecule is excluded from the fitting set when calculating its fitted $\Delta g^\mathrm{OZ/SOC}$, the $c_e$ vary slightly for each molecule. However, the standard deviations are generally small, with relative standard deviations for the chalcogens below 1 \%. Standard deviations are largest for the carbon coefficients, which being smallest suffer most from limited numerical precision in the output data. Notably, the difference in chalcogen-chalcogen interactions between the single- and dual chalcogen pair molecules appears as a general increase in the $c_e$ for the dual pair molecules.

In summary, the linear Ansatz of Eq.~\ref{eq:linmodel} is both highly successful in quantitatively describing the dependence of {\em g}-shifts on changes in local spin density at high orbital angular momentum atoms, yet sensitive to differences in non-local interactions violating the premises of the model.

\begin{table}
\begin{centering}
\begin{tabular}{c|c|c|c}
\hline 
\hline
{\bf Element} & {\bf Avg.} $c_e$ & $c_e$ {\bf SD}  & {\bf RSD [\%]} \\
\multicolumn{4}{c}{Single chalcogen pair molecules} \\
\hline
C & $1.28 \cdot 10^{-4}$ & $1.59 \cdot 10^{-5}$ & 12 \\
S & $7.76 \cdot 10^{-3}$ & $5.63 \cdot 10^{-5}$ & 0 \\
Se & $3.35 \cdot 10^{-2}$ & $1.16 \cdot 10^{-4}$ & 0 \\
\multicolumn{4}{c}{Dual chalcogen pair molecules} \\
\hline
C & $3.06 \cdot 10^{-4}$ & $1.66 \cdot 10^{-5}$ & 5 \\
S & $1.03 \cdot 10^{-2}$ & $1.08 \cdot 10^{-4}$ & 1 \\
Se & $4.66 \cdot 10^{-2}$ & $1.17 \cdot 10^{-4}$ & 0 \\
\end{tabular}
\par\end{centering}

\caption{\label{tab:ce} Average $c_e$ coefficients of Eq.~\ref{eq:linmodel} obtained in the fit of Fig.~\ref{fig:fig3}, and the corresponding standard deviation (SD) and relative standard deviation (RSD) of $c_e$ over the fitting set (see text).}
\end{table}

\subsection{\label{sub:alk} Alkylation}

\begin{figure}
\includegraphics[width=\columnwidth]{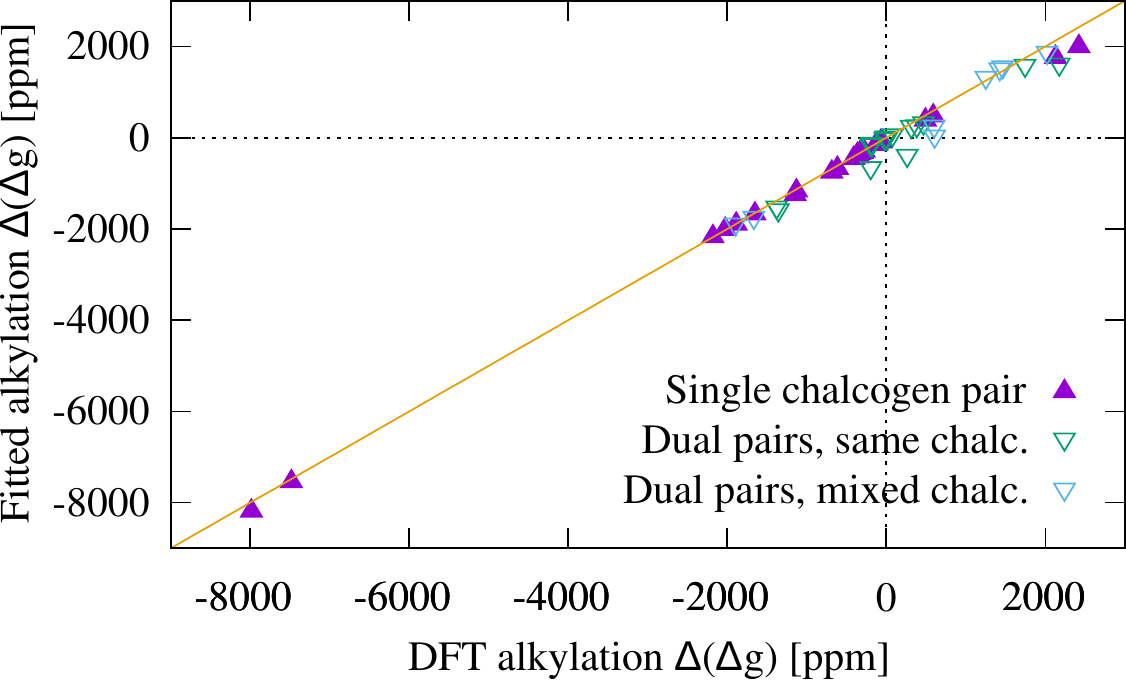}
\caption{\label{fig:fig4} Correlation plot of a fit of the shift in $\Delta g$ upon alkylation of molecules against the corresponding change in local chalcogen spin population, analogous to Eq.~\ref{eq:linmodel} and Fig.~\ref{fig:fig3}. See Fig.~\ref{fig:fig3} for the plot legend. $R^2$ of the entire statistic shown is 0.988.}
\end{figure}

As already briefly discussed above, the effect on predicted {\em g}-shifts of functionalization with an alkyl chain is particularly dramatic.  Alkylation shifts vary from small to negligible shifts independent of alkyl chain bonding site in e.g.~ DATT, DASS, L-DTBTBT and L-DSBSBS, to shifts in on the order of the {\em g}-shift of the pure molecule, either positive or negative depending on the alkyl chain bonding site, as in e.g.~C-DTBTBT or C-DSBTBT. The expectation of merely improving structural properties or solubility of these molecules by adding alkyl chains clearly does not hold here.

In order to elucidate the mechanism behind the alkylation shift, and showcase the utility of our linear Ansatz, we reapply the model to the alkylation shifts analogous to Eq.~\ref{eq:linmodel}.  In Fig.~\ref{fig:fig4}, we have fitted the change in $\Delta g$ (i.e., $\Delta(\Delta g)$) upon alkylation as a linear function of the corresponding change in local spin population at the chalcogen atoms only. The differences to the fit to $\Delta g$ itself in Fig.~\ref{fig:fig3}, is that we do not fit single- and dual-pair molecules separately, nor do we consider changes in spin populations at carbon atoms.

Despite these simplifications, our Ansatz works exceedingly well for the alkylation shifts, with an $R^2$ of the (entire) fit in Fig.~\ref{fig:fig3} of 0.988. This high correlation allows us to identify the mechanism of the alkylation shift as the alkyl chain pushing or pulling spin density onto the chalcogen atoms. The validity of this statement is notably independent of number or composition of the chalcogen atom pairs, geometry, charge confinement etc. Furthermore, and counter-intuitively, it shows how a) an otherwise relatively chemically inert functional group can modulate the effective molecular SOC by bonding site alone, and therefore b) that functional groups may work as a key design element in the tuning of {\em g}-tensors in these and similar OSC molecules.

\subsection{\label{sub:clt} Charge-State Life-Time}

As already touched upon in subsection \ref{sub:dftg}, the consistently lower {\em g}-shifts in the long charge life-time limit (relaxed cationic molecular geometries) may explain discrepancies in the error to experiment of the {\em g}-shifts predicted by DFT. So far, we have not addressed the mechanism behind the suppression of $\Delta g$ at long charge life-time, however. 

Given the analysis above, we already know that strong suppression of $\Delta g$ must be synonymous with spin density depletion at chalcogens. While the $\Delta g$ suppression is strongest in some of the selenium-based molecules, the magnitudes of the effect are not consistent with a mere uniform scaling factor when comparing to the sulfur-based molecules. In fact, the reduction in spin density upon cation structural relaxation of all molecules studied here is roughly twice as large at selenium as at sulfur.

This phenomenon can be understood using concepts from push-pull chemistry, specifically intramolecular charge transfer (ICT) theory. As the geometry of a charged molecule changes to accommodate that charge, the charge equilibrates along molecular bonds by ICT. We may understand the charge balance along these bonds by comparing them to diatomic molecules of elements A and B. In the so-called electronegativity equalization approximation\cite{YongLee2001,Dinur1994} (EEA), the equilibrium charge $Q_{AB}$ due to ICT in such a molecule can be approximated as

\begin{equation}
Q_{AB} = \frac{\Delta \chi_{AB} \cdot \alpha_{AB}}{R^2_{AB}}\,, \label{eq:EEA}
\end{equation}
where $\Delta \chi_{AB}$ is the difference in electronegativity (EN) between A and B, $\alpha_{AB}$ the polarizability of AB, and $R_{AB}$ the interatomic bond length. Eq.~\ref{eq:EEA} allows us to analyse $Q_{AB}$ in terms of estimates of atomic EN and polarizability parameters. Since these parameters effectively depend on the (covalent) bonding state, such an analysis is not strictly valid. Additionally, one should be aware of the limitations of a comparison between a diatomic molecule and an internal bond in a larger molecule.

Still, on a qualitative level, the EEA is useful here. In pure hydrocarbons, ICT will only occur between carbon atoms of similar EN and polarizability. A change in local spin populations upon cationic relaxation in e.g.~pentacene and rubrene therefore mostly depends on changes in bond lengths $R_{AB}$. In the hydrocarbons studied here, such changes are small. Consequentially, so is the suppression of $\Delta g$.

In the chalcogenophene molecules, the change in chalcogen spin populations will depend on the magnitude of the local spin before structural relaxation, the change in bond lengths, and the relative difference in EN and polarizability between the chalcogens and carbon. Carbon, sulfur and selenium atoms all have similar EN of $\chi_\mathrm{C} = 2.55$, $\chi_\mathrm{S} = 2.58$, and $\chi_{\mathrm{Se}} = 2.55$ [Pauling], respectively.\cite{Huheey1993} However, their polarizabilities differ significantly, at $\alpha_\mathrm{C}= 1.76$, $\alpha_\mathrm{S} = 2.90$, and $\alpha_{\mathrm{Se}} = 3.77$ [{\AA}$^3$], respectively.\cite{Politzer2002}

The greater polarizability of the chalcogens pushes positive charge towards the surrounding carbon atoms, and vice versa for negative charge. Since the charge carries the spin, $\Delta g$ upon cation structural relaxation becomes negative for all chalcogen containing molecules. The corresponding changes in carbon - chalcogen bond lengths in the sulfur-based molecules are nearly identical to those in their selenium-substituted analogues. Therefore, the greater spin depletion at selenium atoms evidenced by the stronger $\Delta g$ suppression, can only be explained by the larger polarizability of selenium compared to sulfur. 

These results highlight how the ICT properties of various substituents in conjunction with the charge life-time properties of a given OSC can be used to tune the {\em g}-tensor of the same.

\section{\label{sec:summary}Summary}
Arguably, the strongest argument for the use of OSCs in spintronic applications is their great potential for relatively easy tuning for a specific purpose. We have studied a class of high-mobility chalcogenophene OSCs based on a simple structural theme of phenyl- and chalcogenophene rings functionalized with alkyl chains using density functional theory. Our results show dramatic variations - synonymous with potential tunability - of the molecular {\em g}-tensor shift $\Delta g$ with changes in molecular geometry, extent of the conjugated moiety, chalcogen weight, alkyl chain bonding and charge life-time when ionized. With a single exception, our predictions match available experimental data very well.

Our calculations show that $\Delta g$ in this OSC class is almost entirely determined by the molecular SOC. We analyze our results using a model assuming a linear dependence of the effective molecular SOC on local atomic spin populations. This model accurately reproduces DFT $\Delta g$ when fitted to the same, and therefore explains variations in $\Delta g$ in terms of the overlap of molecular spin- and orbital angular momentum distributions - in other words, the effective SOC. We apply our model to explain the large and counter-intuitive variations in $\Delta g$ with alkyl chain bonding, as well as the effect of structural relaxation of the molecules when positively charged.

In general, this work presents a methodological recipe for a first-principles theoretical analysis of aspects of molecular SOC: the concept of  describing the molecular SOC as dependent on localized atomic spin populations can be applied to numerous phenomena beyond {\em g}-tensor shifts. In particular, our work exhaustively explains how and to what degree tuning of the {\em g}-tensor in this class of OSCs is possible.

\section*{Acknowledgments}
Funding from the Alexander von Humboldt Foundation, the ERC Synergy Grant SC2 (No. 610115), the Transregional Collaborative Research Center (SFB/TRR) 173 SPIN+X, and Grant Agency of the Czech Republic grant no.~14-37427G is acknowledged.

\bibliographystyle{apsrev4-1}
\bibliography{bibliography}

\end{document}